\documentstyle[11pt,aaspp]{article}

\newcommand{\vvec}{{\bf v}}
\newcommand{\xvec}{{\bf x}}
\newcommand{\evec}{{\bf \delta}}
\newcommand{\kvec}{{\bf k}}
\newbox\grsign \setbox\grsign=\hbox{$>$} \newdimen\grdimen \grdimen=\ht\grsign
\newbox\simlessbox \newbox\simgreatbox
\setbox\simgreatbox=\hbox{\raise.5ex\hbox{$>$}\llap
     {\lower.5ex\hbox{$\sim$}}}\ht1=\grdimen\dp1=0pt
\setbox\simlessbox=\hbox{\raise.5ex\hbox{$<$}\llap
     {\lower.5ex\hbox{$\sim$}}}\ht2=\grdimen\dp2=0pt

\begin{document}

\hskip 4in OSU-TA-22/96

\title{Skewness of the Large-Scale Velocity Divergence from Non-Gaussian 
Initial Conditions.}

\author{Zacharias A.M. Protogeros$^{1}$ and Robert J. Scherrer$^{1,2}$}
\affil{E-mail: zack@mps.ohio-state.edu, scherrer@mps.ohio-state.edu}

\footnotetext[1]
{Department of Physics, The Ohio State University,  Columbus, OH 43210}
\footnotetext[2]
{Department of Astronomy, The Ohio State University, Columbus, OH 43210}

\begin{abstract}
We compute the skewness $t_3$ and the corresponding
hierarchical amplitude $T_3$
of the divergence of the velocity field for arbitrary
non-Gaussian initial conditions.  We find that $T_3$ qualitatively
resembles the corresponding hierarchical amplitude for the density field, $S_3$,
in that it contains a term proportional to the initial skewness, which
decays inversely as the linear growth factor, plus a constant
term which differs from the corresponding Gaussian term by a complex
function of the initial three- and four- point functions.  We extend
the results for $S_3$ and $T_3$ with non-Gaussian
initial conditions to evolved fields smoothed with a spherical tophat
window function.  We show that certain linear combinations,
namely $S_3 + {1 \over 2} T_3$, $S_3 + T_3$,
and $s_3 + t_3$, lead
to expressions which are much simpler, for non-Gaussian initial
conditions, than $S_3$ and $T_3$ (or $s_3$ and $t_3$) considered separately.
\end{abstract}


\section{Introduction}

The standard cosmogenesis lore attributes the formation of large-scale 
structure to the enhancement of primordial density fluctuations by gravity.
Given that observations on scales larger than $10$ h$^{-1}$Mpc show the 
amplitude of the rms fluctuations to be less than unity, 
one can successfully apply perturbative techniques to follow the evolution
of the initial probability density function (hereafter PDF).
However, a complete analysis of the problem requires some knowledge of the
statistical nature of the initial fluctuations. Simple inflationary 
scenarios naturally produce initial density fluctuations 
characterized by a Gaussian PDF, whereas
models based on cosmic strings, global texture, or inflation models with
multiple scalar fields (Vilenkin 1985; Kofman 1991; 
Gooding et al. 1992) provide fluctuations which are initially 
non-Gaussian.

A great deal of analytic work has been done for Gaussian initial
conditions.  In particular, the hierarchical amplitudes corresponding
to the skewness $S_3$ and the kurtosis $S_4$ have been known
for some time (Peebles 1980; Fry 1984), and more recently
Bernardeau (1992) has provided a formalism for the calculation
of the full hierarchy.  This work was extended to the skewness of
the velocity field by Bernardeau (1994a) and
Bernardeau et al. (1995), and some progress has been made
toward estimating the evolution of the full PDF (Kofman et al. 1994;
Bernardeau \& Kofman 1995; Protogeros \& Scherrer 1996).
However, a firm 
conclusion has not yet been reached as to the nature of initial conditions 
that would generate, after evolution, large scale structures with the same
statistical characteristics as the ones obtained from observations such as
the QDOT survey or the 1.2 Jy IRAS survey (Coles \& Frenk 1991; Bouchet et 
al. 1993). One reason is that although the statistical 
description of the evolved fields has been calculated in detail for
Gaussian initial conditions, as noted above, the same cannot be said 
for non-Gaussian fields.

For the case of non-Gaussian initial
conditions, a general expression for the evolution of the skewness
was derived by Fry \& Scherrer (1994), and for the kurtosis by
Chodorowski \& Bouchet (1996).
Here we extend this earlier work by examining the evolution
of the divergence of the velocity field ($\theta \equiv \nabla \cdot
\vvec /aH$)
for general non-Gaussian
initial conditions.  Specifically, we calculate the normalized skewness
$t_3=\displaystyle{{\left< \theta^3 \right>}\over{\left<\theta^2 
\right>^{3/2}}}$, as well as the corresponding 
hierarchical coefficient $T_3=\displaystyle{{\left< \theta^3 \right>}\over
{\left<\theta^2 \right>^2}}$ for arbitrary non-Gaussian initial 
conditions. 
(Throughout the paper, we use $S$ to denote statistical indicators related to 
the density field and $T$ to denote the ones related to the 
divergence of the velocity field.)  As noted above, the skewness
of $\theta$ for Gaussian initial conditions has already been
calculated (Bernardeau 1994a; Bernardeau et al. 1995; Bernardeau et al. 1996), as well
as the correlation between $\delta$ and $\theta$ in the quasi-linear
regime (Chodorowski \& Lokas 1996).
       
The calculation of $T_3$ and $t_3$ is presented in the next section.
As in the case of the skewness of the evolved density field, we
find three contributions to $T_3$:
i) the initial, linearly-evolved skewness, which decays
as the inverse of the linear perturbation growth rate,
ii) a ``Gaussian" contribution, identical to the value of $T_3$
for Gaussian initial conditions, and constant in time, and
iii) a term, also constant in time,
which is a complex function
of the initial three- and four-point correlations.
The values of
$T_{3}$ and $S_{3}$ for the smoothed final density field are calculated
in Section 3. We discuss our results in Section 4 and show
that certain linear combinations of $T_3$ and $S_3$ give
simpler expressions than each considered separately.

\section{Calculation of $T_3$}

In what follows we adopt Peebles's (1980) notation in a line of argument paralleling 
that of Fry and Scherrer (1994).
To simplify our derivation, we define
$\tilde \theta \equiv H\theta = \nabla \cdot \vvec/a$.
We begin by expanding $\delta$, $\vvec$, and $\tilde \theta$ perturbatively in the form:
\begin{eqnarray}
\delta & = & \delta^{(1)}+\delta^{(2)}+...\nonumber \\
\vvec & = & \vvec^{(1)}+\vvec^{(2)}+... \nonumber \\
\tilde \theta & = & \tilde \theta^{(1)}+\tilde \theta^{(2)}+...
\end{eqnarray}
so that the skewness of $\tilde \theta$ is given by
\begin{equation}
\label{skewtheta}
\zeta_{\tilde \theta}(0)=\left<\tilde \theta^3 \right>= 
\left<\left[\tilde \theta^{(1)}\right]^3\right> +
3\left< \left[\tilde \theta^{(1)} \right]^2 \tilde\theta^{(2)} \right>+... . 
\end{equation}
The first-order and second-order solutions for $\delta$ are the well-known
linear result:
\begin{equation}
\delta^{(1)} = D_1(t) \delta(\xvec, t_0),
\end{equation}
where $D_1(t)$ is the usual growing mode solution,
and (Peebles 1980),
\begin{equation}
\delta^{(2)} = {5\over7}\delta_0^2-\delta_{0,i} 
\Delta_{0,i}+{2\over 7}\Delta_{0,ij}\Delta_{0,ij},
\end{equation}
where we have defined $\delta_0 \equiv \delta^{(1)}$,
and $\Delta_0$
is defined to be:
\begin{equation}
\Delta_{0}={1\over{4\pi}} \int
d^3x'\delta_0(\xvec'){1\over {\vert \xvec-\xvec'\vert}}.
\end{equation} 

To go from these expressions to the values for
$\tilde \theta^{(2)}= \nabla\cdot\vvec^{(2)}/a$ and
$\tilde \theta^{(1)}=\nabla\cdot\vvec^{(1)}/a$,
we substitute the expansions for $\delta$ and $\vvec$
into the continuity equation:
\begin{equation}
\dot\delta + {1\over a}\nabla\cdot(1+\delta)\vvec = 0,
\end{equation}
leading to the linear and second-order equations:
\begin{eqnarray}
\dot\delta^{(1)}+{1\over a}\nabla\cdot \vvec^{(1)} & = & 0, \\
\dot\delta^{(2)}+{1\over a}\nabla \cdot \vvec^{(2)}+ {1\over a}
\delta^{(1)}\nabla\cdot \vvec^{(1)}+{1\over a}
\vvec^{(1)}\cdot\nabla\delta^{(1)} & = & 0.
\end{eqnarray}
The first-order equation gives
$v^{(1)}_{i} = a\dot D_1 \Delta_{0,i}$ and $\nabla\cdot\vvec^{(1)} = 
-a\dot D_1 \delta_0(\xvec)$, so that
\begin{equation}
\label{theta1}
\tilde \theta^{(1)} = -\dot D_1 \delta_0(\xvec),
\end{equation}
while the second-order equation gives:
\begin{equation}
\label{theta2}
\tilde \theta^{(2)}=-D_1\dot D_1 \left({3\over 7}\delta_0^2-\delta_{0,i}
\Delta_{0,i}+
{4\over 7}\Delta_{0,ij}\Delta_{0,ij} \right).
\end{equation}
From equations (\ref{theta1}) and (\ref{theta2}),
we obtain:
\begin{equation}
\label{theta3}
\left<\left[\tilde \theta^{(1)}\right]^3\right> =
-\dot D_1^3 \left< \left[ \delta_0 \right]^3 \right>,
\end{equation}
and
\begin{equation}
\label{skewtheta2}
\left< \left[\tilde \theta^{(1)} \right]^2 \tilde\theta^{(2)} \right> =
-\dot D_1^3 D_1 \left[{3\over7} \left<\delta_0^4 \right> - 
\left<\delta_0^2 \delta_{0,i} \Delta_{0,i} \right>
 + {4\over 7} \left<\delta_0^2 \Delta_{0,ij}\Delta_{0,ij} 
\right> \right].
\end{equation}
We now simplify the various terms in equation (\ref{skewtheta2}) using the results
of Peebles (1980) and Fry \& Scherrer (1994).

The second-, third-, and fourth-order moments
of the density field can be expressed in terms of the
irreducible 2-,3- and 4-point correlation functions
$\xi_{12}$, $\zeta_{123}$, and $\eta_{1234}$:
\begin{eqnarray}
\left<\delta(\xvec_{1})\delta(\xvec_{2}) \right> & = & \xi_{12}, \nonumber \\
\left<\delta(\xvec_{1})\delta(\xvec_{2})\delta(\xvec_{3}) \right> & = & 
\zeta_{123}, \nonumber \\
\left<\delta(\xvec_{1})\delta(\xvec_{2})\delta(\xvec_{3})\delta(\xvec_{4})
\right> & = & \xi_{12}\xi_{34}+\xi_{13}\xi_{24}+\xi_{23}\xi_{14}+\eta_{1234}.
\end{eqnarray}
where all the moments are functions of $\vert \xvec-\xvec' \vert$ due to the 
assumed homogeneity and isotropy of the density field distribution. Therefore, 
\begin{equation}
\label{delta4}
\left<\delta_0^4 \right> =  3 \xi_0^2(0) 
+\eta_0(0),
\end{equation}
and
\begin{eqnarray}
\label{del3Del}
\left<\delta_0^2\delta_{0,i}\Delta_{0,i} \right> & = & 
{1\over {4\pi}}\int d^3x'{1\over {\vert \xvec - \xvec' \vert}},_{i}
\xi_0(0)\xi_0(\xvec-\xvec')_{,i}\nonumber\\
& + & {1\over 3}{1\over {4\pi}} \int d^3x'
{1\over {\vert \xvec-\xvec' \vert}},_{i}
\eta_0(\xvec,\xvec,\xvec,\xvec')_{,i}, 
\end{eqnarray}
where the zero subscripts indicate linearly-evolved quantities.
The first term in equation (\ref{del3Del})
yields $\xi_0^2(0)$, whereas the second term yields ${1\over 3}
\eta_0(0)$ upon integration by parts, using the fact that
$\eta(\xvec,\xvec,\xvec,\xvec)=\eta(0)$ from isotropy.
Thus,
\begin{equation}
\label{delta3Delta}
\left<\delta_0^2 \delta_{0,i} \Delta_{0,i} \right> = 
\xi_0^2(0)+ {1\over 3}\eta_0(0).
\end{equation}  
Finally, we evaluate the last term in equation (\ref{skewtheta2}):
\begin{eqnarray}
\label{del2Del2}
 \left<\delta_0^2\Delta_{0,ij}\Delta_{0,ij} \right>  =  
{1\over {(4\pi)^2}} \int\int d^3x' d^3x'' {1\over {\vert \xvec-\xvec'\vert}},_
{ij}{1\over {\vert \xvec-\xvec''\vert}},_{ij} 
\left<\delta_0^2(\xvec)\delta_0(\xvec')\delta_0(\xvec'') \right>.
\end{eqnarray}
The fourth moment on the right-hand side can be expressed as
\begin{equation}
\left<\delta_0^2(\xvec)\delta_0(\xvec')\delta_0(\xvec'') \right> =
\xi_0(0)\xi_0(\xvec'-\xvec'') +
2\xi_0(\xvec-\xvec')\xi_0(\xvec-\xvec'')+\eta_0(\xvec,\xvec,\xvec',\xvec'').
\end{equation}
The first two terms in the double integration in equation (\ref{del2Del2})
yield $\xi_0^2(0)$ and
${2\over 3} \xi_0^2(0)$ respectively, for a total contribution of ${5\over 3}
\xi_0^2(0)$. The last term yields $I[\eta_0] + {1\over 3} \eta_0(0)$, where we
define (Fry \& Scherrer 1994):
\begin{equation}
\label{Iintegral}
I[\eta_0]\equiv{1\over {(4\pi)^2}}\int\int d^3x'd^3x''\eta_0(0,0,\xvec',\xvec'')
{6P_2(\hat x' \cdot \hat x'') \over {x'^3x''^3}},
\end{equation}
with $P_2(\hat x' \cdot \hat x'')$ the second Legendre polynomial.
Therefore,
\begin{equation}
\label{delta2Delta2}
\left<\delta_0^2\Delta_{0,ij}\Delta_{0,ij} \right>={5\over
3}\xi_0^2(0)+I[\eta_0]+{1\over 3}\eta_0(0).
\end{equation}
Combining the results of equations (\ref{delta4}), (\ref{delta3Delta}),
and (\ref{delta2Delta2}), equation
(\ref{skewtheta2}) becomes
\begin{equation}
\label{finaltheta2}
\left< \left[\tilde \theta^{(1)} \right]^2 \tilde\theta^{(2)} \right>=
-\dot D_1^3 D_1 \left[{26\over{21}}\xi_0^2(0)+{6\over{21}}\eta_0(0)+{4 \over 7}
I[\eta_0] \right].
\end{equation}
Substituting equations (\ref{theta3}) and (\ref{finaltheta2}) into
equation (\ref{skewtheta}), we obtain our final expression
for $\zeta_{\tilde\theta}(0)$:
\begin{equation}
\label{final1}
\zeta_{\tilde\theta}(0)=-\dot D_1^3 D_1 \left[{ {\zeta_0(0) \over {D_1}} + 
{26 \over {7}}\xi_0^2(0) + {6 \over {7}}\eta_0(0) + {12 \over {7}}I[\eta_0] }
\right].
\end{equation}

To calculate the hierarchical amplitude $T_3$ or the normalized
skewness $t_3$, we must also calculate
$\xi_{ \tilde \theta}(0) \equiv \left<\tilde \theta^2 \right>$, which
can be derived in a calculation similar to that for $\zeta_{\tilde \theta}$.
We have
\begin{equation}
\xi_{ \tilde \theta} = \left<\tilde \theta^2 \right> = 
\left< \left[\tilde\theta^{(1)} \right]^2 \right>+2 \left<\tilde\theta^{(1)}
\tilde\theta^{(2)} \right> + ...
\end{equation}
with
\begin{equation}
\left< \left[\tilde\theta^{(1)} \right]^2 \right>=\dot D_1^2 \xi_0(0)
\end{equation}
and
\begin{equation}
\label{theta1theta2}
\left<\tilde\theta^{(1)}\tilde\theta^{(2)} \right>=\dot D_1^2 D_1 \left<{3\over
{7}}\delta_0^3-\delta_0\delta_{0,i}\Delta_{0,i} + {4\over
{7}}\delta_0\Delta_{0,ij}\Delta_{0,ij} \right>.
\end{equation}
In a manner similar to our previous derivations, we obtain
\begin{eqnarray}
\left<\delta_0^3 \right>&=&\zeta_0(0), \\
\left<\delta_0\delta_{0,i}\Delta_{0,i} \right>&=&{1\over {2}} 
\zeta_0(0), \\
\left<\delta_0\Delta_{0,ij}\Delta_{0,ij} \right>&=&
I[\zeta_0]+{1\over{3}}\zeta_0(0),
\end{eqnarray}
where $I[\zeta_0]$ is the integral analogous to equation (\ref{Iintegral}),
integrated over $\zeta_0(0,x',x'')$, rather than $\eta_0(0,0,x',x'')$.
Therefore, $\left<\tilde\theta^{(1)}\tilde\theta^{(2)} \right>=
\displaystyle {1\over {2}} \left[{5\over{21}}\zeta_0(0)+{8\over {7}}
I[\zeta_0] \right]$ and
\begin{equation}
\label{final2}
\xi_{\tilde\theta}(0)=\left<\tilde\theta^2 \right>=
\dot D_1^2 D_1 \left[{\xi_0(0) \over {D_1}} + {5\over{21}}\zeta_0(0)+
{8\over {7}}I[\zeta_0] \right].
\end{equation}

Combining equations (\ref{final1}) and (\ref{final2}), with
$\theta= \displaystyle{1\over {H}} \tilde\theta$, and defining
$f(\Omega)=\displaystyle{1\over H}{\dot D_1 \over D_1}\approx
\Omega^{0.6}$,
we obtain our
expressions $T_3$ and $t_3$ for arbitrary non-Gaussian initial conditions.
For the hierarchical amplitude $T_3$, we get
\begin{equation}
\label{T3}
T_3=-{1\over {f(\Omega)}} \left[{1\over {D_1(t)}}{\zeta_0(0)\over {\xi_0^2(0)}} 
+{26 \over
{7}} + {6\over {7}}{\eta_0(0) \over {\xi_0^2(0)}} + {12\over {7}}{I[\eta_0]
\over{\xi_0^2(0)}}-{10\over {21}}{\zeta_0^2(0) \over{\xi_0^3(0)}}-{16\over
{7}}{\zeta_0(0)I[\zeta_0] \over {\xi_0^3(0)}} \right],
\end{equation}
where only terms of order up to $O(\sigma^0)$ have been retained.
In the case of Gaussian initial conditions, all of the terms except the second
term are zero, yielding the standard result (Bernardeau 1994a,
Bernardeau et al. 1995)
\begin{equation}
T_3^G = - {26 \over 7} \Omega^{-0.6}.
\end{equation}
For comparison the hierarchical coefficient for the density field
is (Fry \& Scherrer 1994)
\begin{equation}
\label{S3}
S_3  =  {1\over {D_1}}{\zeta_0(0)\over {\xi_0^2(0)}} 
+{34 \over{7}} + {10\over {7}}{\eta_0(0) \over {\xi_0^2(0)}} + 
{6\over {7}}{I[\eta_0]\over{\xi_0^2(0)}}-{26\over {21}}{\zeta_0^2(0) 
\over{\xi_0^3(0)}}-{8\over{7}}{\zeta_0(0)I[\zeta_0] \over {\xi_0^3(0)}}.
\end{equation}
We see that $T_3$, like $S_3$, contains three distinct contributions:  a term
proportional to the initial skewness which decays away as $1/D_1(t)$, a
``Gaussian" piece which is constant and identical to the hierarchical
amplitude in the Gaussian case, and a third contribution, also constant in time,
which is a complex function of the initial skewness, the initial kurtosis, and
various integrals over the initial three- and four-point functions.
In fact, a comparison of equations (\ref{S3}) and (\ref{T3}) indicates that
the terms in the two equations are identical functions of the initial
density field; only the coefficients multiplying the various terms
are different.  We will exploit this fact in Section 4.

For the normalized skewness $t_3$ we obtain
\begin{eqnarray}
t_3 & = & -s_{3,0} \nonumber \\
    & - & \left[ {26\over 7} + {6\over 7}{{\eta_0(0)}\over {\xi_0^2(0)}}+
          {12\over 7}{{I[\eta_0]}\over {\xi_0^2(0)}}
      -{5\over {14}}{{\zeta_0^2(0)}\over {\xi_0^3(0)}}-{12\over 7}
          {{\zeta_0(0)I[\zeta_0]}\over{\xi_0^3(0)}} \right]\sigma_0(t),
\end{eqnarray}
which can be compared with the corresponding normalized skewness for
the density (Fry \& Scherrer 1994)
\begin{eqnarray}
s_3 & = & s_{3,0}\nonumber \\
    & + & \left[{34\over 7}+{10\over 7}{\eta_0(0) \over {\xi_0^2(0)}}+
{6\over 7}{{I[\eta_0]}\over {\xi_0^2(0)}}-{13\over {14}}{{\zeta_0^2(0)}
\over {\xi_0^3(0)}}-{6\over 7}{{\zeta_0(0)I[\zeta_0]}\over{\xi_0^3(0)}} 
\right]\sigma_0(t).
\end{eqnarray}
In these expressions, $\sigma_0(t)$ is the linearly-evolved
rms fluctuation:  $\sigma_0^2(t) = \xi_0(0)$, and $s_{3,0}$ is
the (constant) linearly-evolved skewness:  $s_{3,0} = 
\zeta_0(0) / \sigma_0^3(t)$.  Note that $t_3$, unlike
$T_3$, is independent of $f(\Omega)$.

\section{Smoothed $T_3$ and $S_3$ results}

Although the results derived in the previous section are interesting
from a formal point of view, an application to the observations
requires a calculation of $T_3$ for a field which has been smoothed
with a window function.  Both Fry \& Scherrer (1994) and
Chodorowski \& Bouchet (1996) argued that the smoothed skewness
and kurtosis for non-Gaussian initial conditions should qualitatively resemble the unsmoothed results.
The effects of smoothing on the hierarchical
amplitudes for Gaussian initial conditions have been calculated in
detail for the case of spherical tophat smoothing (Bernardeau 1994a,b),
so we will follow Bernardeau's treatment to derive an expression
for the value of $T_3$ with non-Gaussian initial conditions and spherical
tophat smoothing.  In addition, we extend the results of
Fry \& Scherrer (1994) by performing the same calculation to derive
the smoothed value of $S_3$.

Consider a spherical tophat window function with radius $R_0$ and Fourier
transform
\begin{equation}
W(kR_0)={2\over {(kR_0)^3}}\left[\sin(kR_0)-kR_0\cos(kR_0) \right].
\end{equation}
After the density and velocity-divergence fields have been smoothed
with this window function, we obtain new expressions for
$S_3$ and $T_3$, which we denote by $S_3(R_0)$ and $T_3(R_0)$.
We now use the methods developed in Bernardeau (1994a) to calculate
these quantities.
The expressions for the smoothed $\theta$ up to second order are
(Bernardeau 1994a)
\begin{eqnarray}
\theta^{(1)}(R_0) & = & -f(\Omega)D_1 \int {d^3k \over {(2\pi)^{3/2}}}\evec_{\kvec}
W(kR_0) \nonumber \\ 
\theta^{(2)}(R_0) & = & - f(\Omega) \int  {d^3k_1 \over {(2\pi)^{3/2}}} {d^3k_2 \over 
{(2\pi)^{3/2}}}\evec_{\kvec_1}\evec_{\kvec_2}W(|\kvec_1+\kvec_2|R_0) 
 \left[D_1^2(P_{12}-{3\over 2}Q_{12})+{3\over 4}E_2Q_{12} \right]
\end{eqnarray}
where $\delta_\kvec$ is the Fourier transform of the initial density field, 
the quantities $P_{ij}$, 
$Q_{ij}$ are defined by
\begin{eqnarray}
P_{ij} & = & 1+{{\kvec_i \cdot \kvec_j}\over {k_i^2}}, \nonumber \\
Q_{ij} & = & 1-{{(\kvec_i \cdot \kvec_j)^2}\over {k_i^2k_j^2}},
\end{eqnarray}
and
$D_i$ gives the time dependence of the growing mode, being the 
solution of the $i$th order time evolution equation for $\delta$, while
$E_2=\displaystyle D_1{d \over {dD_1}}(D_2-D_1^2)$. At the limit of 
$t \rightarrow 0$, $D_2$ satisfies the relation 
$ D_2(t)\approx \displaystyle{34\over 21}D_1^2(t)$.
Furthermore, the 2,3 and 4-point functions of the random field 
$\delta_{\kvec}$ are expressed as
\begin{eqnarray}
\left < \evec_{\kvec_1}\evec_{\kvec_2} \right> & = & \xi_{\kvec_{12}}, \\
\left < \evec_{\kvec_1}\evec_{\kvec_2} \evec_{\kvec_3} \right> & = & \zeta_{
\kvec_{123}}, \\
\left < \evec_{\kvec_1}\evec_{\kvec_2}\evec_{\kvec_3}\evec_{\kvec_4} \right> & = 
& \xi_{\kvec_{12}}\xi_{\kvec_{34}}+\xi_{\kvec_{13}}\xi_{\kvec_{24}}+
\xi_{\kvec_{23}}\xi_{\kvec_{14}}+\eta_{\kvec_{1234}},
\end{eqnarray}
where the 2-point function is related to the power spectrum $P(k)$ of the 
$\delta_{\kvec}$ field as $\xi_{\kvec_{ij}}=\delta_D(\kvec_i+\kvec_j)P(k_i)$.
Using these relations and given that $\left<\theta^2 \right> = \left< 
\left[\theta^{(1)} \right]^2 \right>+2 \left<\theta^{(1)} \theta^{(2)} 
\right> + ...,$  
we obtain
\begin{equation}
\label{theta2R}
\left < \theta^2(R_0) \right> = f(\Omega)^2 D_1^2 \sigma^2(R_0)+
2 f(\Omega)^2D_1^3 I_1 \left[ \zeta_{\kvec_{123}} \right]
\end{equation}
where we have used the definitions
\begin{equation}
\sigma^2(R_0)=\int {d^3k \over {(2\pi)^{3}}}W(kR_0)^2P(k),
\end{equation}
and
\begin{equation}
I_1 \left[ \zeta_{\kvec_{123}} \right]=\int {d^3k_1 \over {(2\pi)^{3/2}}}
{d^3k_2 \over {(2\pi)^{3/2}}}{d^3k_3 \over {(2\pi)^{3/2}}}W_1W_{23}\zeta_
{\kvec_{123}} \left[(P_{23}-{3\over 2}Q_{23})+ {3\over 4}{E_2\over{D_1^2}}
Q_{23} \right],
\end{equation}
where $W_i=W(k_iR_0)$, and $W_{ij}=W(|\kvec_i+\kvec_j|R_0)$.
In a similar fashion, we obtain
\begin{equation}
\left< \left[ \theta^{(1)} \right]^3 \right> = -f(\Omega)^3D_1^3I_2 
\left[ \zeta_{\kvec_{123}} \right],
\end{equation}
where we define
\begin{equation}
I_2 \left[ \zeta_{\kvec_{123}} \right]=\int {d^3k_1 \over {(2\pi)^{3/2}}}
{d^3k_2 \over {(2\pi)^{3/2}}}{d^3k_3 \over {(2\pi)^{3/2}}}W_1W_2W_3\zeta_
{\kvec_{123}}.
\end{equation}
Finally, as shown in the Appendix,  we get
\begin{equation}
\label{appendix}
\left< \left[ \theta^{(1)} \right]^2 \theta^{(2)} \right> = 
-f(\Omega)^3D_1^4\sigma^4(R_0) \left({26\over 21} + 
{\gamma\over 3}\right) - f(\Omega)^3D_1^4I\left[ \eta_{\kvec_{1234}} \right],
\end{equation}
where we define
\begin{equation}
I\left[ \eta_{\kvec_{1234}} \right]=\int {d^3k_1 \over {(2\pi)^{3/2}}}
{d^3k_2 \over {(2\pi)^{3/2}}}{d^3k_3 \over {(2\pi)^{3/2}}}
{d^3k_4 \over {(2\pi)^{3/2}}} W_1W_2W_{34}\eta_
{\kvec_{1234}} \left[(P_{34}-{3\over 2}Q_{34})+ {3\over 4}{E_2\over {D_1^2}}
Q_{34} \right].
\end{equation}
Since $\left < \theta^3 \right> = \left< 
\left[ \theta^{(1)} \right]^3 \right> + 3 \left< \left[ \theta^{(1)} 
\right]^2 \theta^{(2)} \right> + ...$, we obtain
\begin{equation}
\left< \theta^3(R_0) \right> = -f(\Omega)^3D_1^3I_2\left[\zeta_{\kvec_{123}} 
\right] - f(\Omega)^3D_1^4 \sigma^4(R_0) \left({26\over 7} + 
\gamma \right)-3f(\Omega)^3D_1^4I\left[\eta_{\kvec_{1234}}\right],
\end{equation}
where $\gamma$ is defined as
\begin{equation}
\gamma={ {d\log\left[ \sigma^2(R_0) \right]}
\over {d\log(R_0)}}.
\end{equation}
Using these results it is easy to show then that
\begin{eqnarray}
T_3(R_0) & = & -{1\over f(\Omega)} { {I_2\left[\zeta_{\kvec_{123}} \right]}\over 
{D_1\sigma^4(R_0)}}\nonumber \\
       & - & {3\over f(\Omega)} {{I\left[\eta_{\kvec_{1234}} \right]}\over 
{\sigma^4(R_0)}} + {4\over f(\Omega)} {{I_1\left
[\zeta_{\kvec_{123}} \right]I_2\left[\zeta_{\kvec_{123}} \right]}
\over {\sigma^6(R_0)}} - 
{1\over f(\Omega)}\left({26\over 7}+\gamma \right). \\
\end{eqnarray}
As in the unsmoothed case, the first term decays as $1/D_1(t)$, while
the rest of the expression is constant in time.
Observe that for Gaussian initial conditions our result reduces to
\begin{eqnarray}
T^G_3(R_0) & = &-{1\over f(\Omega)}\left( {26\over 7}+ \gamma 
\right),\nonumber \\
	 & = &-{1\over {\Omega^{0.6}}}\left[ {26\over 7}-(n+3) \right],
\end{eqnarray}
in agreement with Bernardeau (1994a),
where the last relation holds for a power spectrum 
$P(k)\propto k^n$. 

Using the same line of argument, it is easy to show that
for the smoothed density field, the mean values of $\delta^2$
and $\delta^3$ are
\begin{equation}
\label{delta2R}
\left < \delta^2(R_0) \right > = D_1^2 \sigma^2(R_0)+
2 D_1^3 \left [I_1 \left[ \zeta_{\kvec_{123}} \right] + K
\left [ \zeta_{\kvec_{123}} \right ] \right ],
\end{equation}
and
\begin{equation}
\left < \delta^3(R_0) \right > = D_1^3 I_2\left[\zeta_{\kvec_{123}} 
\right] + D_1^4 \sigma^4(R_0) \left({34\over 7} + 
\gamma \right) + 3 D_1^4 \left [ I\left[\eta_{\kvec_{1234}}\right] + K\left[\eta_{\kvec_{1234}}\right] \right ].
\end{equation}
Then the skewness of the density field smoothed with a spherical
tophat window function for
non-Gaussian initial conditions is
\begin{eqnarray}
S_3(R_0) & = & { {I_2\left[\zeta_{\kvec_{123}} \right]}\over 
{D_1\sigma^4(R_0)}}\nonumber \\
       & + &\left({34\over 7}+\gamma \right)+{3\over {\sigma^4(R_0)}} 
	\left[I\left[\eta_{\kvec_{1234}} \right]+K\left[\eta_{\kvec_{1234}}
        \right]\right]\nonumber \\
       & - & {4\over{\sigma^6(R_0)}}\left[I_1\left[\zeta_{\kvec_{123}}
\right]I_2\left[\zeta_{\kvec_{123}}\right]+I_2\left[\zeta_{\kvec_{123}}\right]
K\left[\zeta_{\kvec_{123}}\right] \right],
\end{eqnarray}
where we use the definitions
\begin{eqnarray}
K\left[\zeta_{\kvec_{123}}\right] & = & {2\over 7}
\int\int\int {d^3k_1 \over 
{(2\pi)^{3/2}}}{d^3k_2 \over {(2\pi)^{3/2}}}{d^3k_3 \over {(2\pi)^{3/2}}} 
Q_{23}W_{1}W_{23}\zeta_{\kvec_{123}}\nonumber \\
K\left[\eta_{\kvec_{1234}}\right] & = & {2\over 7}
\int\int\int\int {d^3k_1 \over 
{(2\pi)^{3/2}}}{d^3k_2 \over {(2\pi)^{3/2}}}{d^3k_3 \over {(2\pi)^{3/2}}}
{d^3k_4 \over {(2\pi)^{3/2}}} Q_{34}W_{1}W_{2}W_{34}\eta_{\kvec_{1234}}
\end{eqnarray}
and we have kept terms up to $O(\sigma^0)$.
For Gaussian initial conditions our result reduces to
\begin{eqnarray}
S_3^G(R_0) & = & {34\over 7}+\gamma, \nonumber \\
         & = & {34\over 7}-(n+3),
\end{eqnarray}
for a power spectrum $P(k)\propto k^n$,
consistent with Bernardeau (1994a).

\section{Discussion}
A calculation of this sort can be used in two different ways:
to gain further insight into the process of gravitational
clustering at a fundamental level, and to compare with
observations to try to determine if the initial perturbations
were non-Gaussian.  For the latter purpose, only our smoothed
results are useful, although the unsmoothed results may
be of interest for the former.

As noted earlier, the hierarchical amplitude for the velocity divergence,
$T_3$ resembles qualitatively the corresponding expression
for the density field, $S_3$.  We end up with one term, proportional
to the initial skewness, which decays as $1/D_1$, while the term which
is constant in time can be broken down into a ``Gaussian" piece,
equal to the contribution for Gaussian initial conditions, and a
``non-Gaussian" piece, which depends in a complex manner on
the three- and four-point functions of the initial density field.

A problem with applying the results of either Section 2 or Section 3
is their complexity in comparison with their Gaussian counterparts.
Even the unsmoothed results are non-local, depending on integrals
over the initial distribution functions.  However, it is possible
to simplify these results somewhat by examining combinations
of $S_3$ and $T_3$.  Consider first the unsmoothed case for $\Omega=1$.
In this case, the non-local terms arise from the last term in equation
(\ref{skewtheta2}).  These can be eliminated by evaluating
$S_3 + {1\over2}T_3$, for which we obtain
\begin{equation}
S_3 + {1\over 2}T_3 = {1\over 2D_1}{\zeta_0(0) \over \xi_0^2(0)}
+ 3 + {\eta_0(0) \over {\xi_0^2(0)}}
- {\zeta_0^2(0) 
\over{\xi_0^3(0)}}.
\end{equation}
This expression is a function only of the initial skewness and kurtosis
of the non-Gaussian density field.

A more useful combination, from the point of view of the observations,
can be derived from the smoothed hierarchical amplitudes given
in the previous section.  For the case $\Omega = 1$,
if we simply take the sum of $S_3(R_0)$
and $T_3(R_0)$, we obtain:
\begin{equation}
\label{S3+T3}
S_3(R_0)+T_3(R_0)={8\over 7}+{{3K\left[\eta_{\kvec_{1234}}\right]}\over {\sigma^4(R_0)}}-{{4I_2\left[\zeta_{\kvec_{123}}\right]}
K\left[\zeta_{\kvec_{123}}\right]\over {\sigma^6(R_0)}}.
\end{equation}
This quantity has several interesting properties.  For the case
of Gaussian initial conditions, it reduces to
\begin{equation}
S_3^G(R_0)+T_3^G(R_0)={8\over 7},
\end{equation}
which is independent of the initial power spectrum.  This result does not extend
to higher-order amplitudes; e.g., $S_4 - T_4$ does depend on the initial
power spectrum.  For non-Gaussian initial conditions, the time-dependent
term produced by the initial skewness has vanished,
so equation (\ref{S3+T3}) gives a much cleaner estimate of the
deviation from Gaussian hierarchical clustering for non-Gaussian
initial conditions; any deviation from 8/7 indicates the presence
of non-Gaussian initial conditions or $\Omega \ne 1$.

One of the main reasons for investigating the behavior
of $T_3(R_0)$ is its sensitivity to different values of $\Omega$
(Bernardeau et al. 1995; Bernardeau et al. 1996).
Unfortunately, this works to our disadvantage in
equation (\ref{S3+T3}):  it is not possible to disentangle the effects
of $\Omega \ne 1$ from the effects of non-Gaussian initial
conditions.  A more useful quantity if one is interested in
the statistics of the initial conditions is $s_3(R_0) + t_3(R_0)$, since
$t_3$ is independent of $f(\Omega)$.  A straightforward calculation
similar to our derivation of $S_3(R_0)+T_3(R_0)$
leads to:
\begin{equation}
\label{s3+t3}
s_3(R_0)+t_3(R_0)=\left[{8\over 7}+{{3K\left[\eta_{\kvec_{1234}}\right]}\over {\sigma^4(R_0)}}-{{3I_2\left[\zeta_{\kvec_{123}}\right]}
K\left[\zeta_{\kvec_{123}}\right]\over {\sigma^6(R_0)}}\right] D_1 \sigma(R_0),
\end{equation}
a result which holds for any value of $\Omega$.  Thus, equation
(\ref{s3+t3}) provides a clear distinction between the evolution of
Gaussian and non-Gaussian initial conditions.

The application of hierarchical amplitudes to the case of non-Gaussian
initial conditions continues to be difficult, due primarily
to the much greater complexity of the results.  However, our calculations
provide some simpler expressions which represent a step
in the right direction.

\vskip 1 cm

\centerline {\bf Acknowledgments}

\indent

Z.A.M.P. and R.J.S. were supported in part by the
Department of Energy (DE-AC02-76ER01545).  R.J.S was
supported in part
by NASA (NAG 5-2864 and NAG 5-3111).

\vfill
\eject
\centerline{\bf REFERENCES} 

\vskip 1 cm 

\noindent Bernardeau, F. 1992, ApJ, 392, 1

\noindent Bernardeau, F. 1994a, ApJ, 433, 1

\noindent Bernardeau, F. 1994b, A \& A, 291, 697

\noindent Bernardeau, F., Juszkiewicz, R., Dekel, A., \& Bouchet, F.R.
1995, MNRAS, 274, 20

\noindent Bernardeau, F., \& Kofman, L. 1995, ApJ, 443, 479

\noindent Bernardeau, F., van de Weygaert, R., Hivon, E., \& Bouchet, F.R.
1996, MNRAS, submitted (astro-ph/9609027)

\noindent Bouchet, F.R., et al. 1993, ApJ, 417, 36

\noindent Chodorowski, M.J., \& Bouchet, F.R. 1996, MNRAS, 279, 557

\noindent Chodorowski, M.J., \& Lokas, E.L. 1996, MNRAS, submitted
(astro-ph/9606088)

\noindent Coles, P., \& Frenk, C.S. 1991, MNRAS, 253, 727

\noindent Fry, J.N. 1984, ApJ, 279, 499

\noindent Fry, J.N., \& Scherrer, R.J. 1994, ApJ, 429, 36

\noindent Gooding, A.K., Park, C., Spergel, D.N., Turok, N., \&
Gott, J.R. 1992, ApJ, 393, 42

\noindent Kofman, L. 1991, Phys Scripta, T36, 108

\noindent Kofman, L., Bertschinger, E., Gelb, J.M., Nusser, A., \& Dekel, A.
1994, ApJ, 420, 44

\noindent Peebles, P.J.E. 1980, The Large-Scale Structure of the Universe
(Princeton University Press)

\noindent Protogeros, Z.A.M., \& Scherrer, R.J. 1996, MNRAS, in press
(astro-ph/9603155)

\noindent Vilenkin, A. 1985, Phys Rep, 121, 265

\vfill
\eject

\centerline{\bf Appendix}

In Section 2 we used the integrals
\begin{eqnarray}
{1\over {(4\pi)^2}}\int\int d^3xd^3x'\xi(\xvec'-\xvec''){6P_2(\hat
\xvec'\cdot\hat\xvec'') \over {x'^3x''^3}} & = & {2 \over 3}\xi(0),\\
{1\over {(4\pi)^2}}\int\int d^3xd^3x'\xi(\xvec'-\xvec'')({1 \over {\vert
\xvec-\xvec'\vert}})_{ij}({1 \over {\vert\xvec-\xvec''\vert}})_{ij} & = & \xi(0),
\end{eqnarray}
which are easily derived using an integration by parts and the relation
\begin{eqnarray}
\nabla_i\nabla_j{1\over {|\xvec|}}={ {3\hat\xvec_i \cdot \hat\xvec_j-
\delta_{ij}}\over x^3}-{{4\pi}\over 3}\delta_{ij}\delta_D(\xvec).
\end{eqnarray}
In Section 3 we used the results given by Bernardeau's equations (A26) and (A27) (Bernardeau 1994a), specifically,
\begin{eqnarray}
\int\int {d^3k_1 \over {(2\pi)^{3}}}
{d^3k_2 \over {(2\pi)^{3}}}P_{12}W_{1}W_2W_{12}P(k_1)P(k_2) & = & 
\sigma^4(R_0)\left[ 1 + {1\over 6}{ {R_0}\over 
{\sigma^2(R_0)}}{ {d\sigma^2(R_0)}\over 
{dR_0}} \right], \\ 
\int\int {d^3k_1 \over {(2\pi)^{3}}}
{d^3k_2 \over {(2\pi)^{3}}}Q_{12}W_{1}W_2W_{12}P(k_1)P(k_2) & = & {2\over 3}
\sigma^4(R_0), \\
\end{eqnarray}
and the fact that in the equation for
$\left< \left[ \theta^{(1)} \right]^2 \theta^{(2)} 
\right>$, integrals containing the factors
$P_{ij} \delta(\kvec_i + \kvec_j)$ and
$Q_{ij} \delta(\kvec_i + \kvec_j)$ vanish.

\end{document}